\documentclass[reprint,amsmath]{revtex4-2}
\usepackage{bm}
\usepackage{graphicx}
\usepackage{xcolor}
\usepackage{tikz}

\newcommand{\dz}[1]{\frac{\partial #1}{\partial z}}

\begin{document}

\title{A force model for dense granular flows and particle segregation}

\author{M.P. van Schrojenstein Lantman}
\email{m.v.s.lantman@marin.nl}
\affiliation{Research and Development, Maritime Research Institute Netherlands, Wageningen 6708, the Netherlands}

\date{\today}

\begin{abstract}
Dense granular flows are a common occurrence throughout nature and industry, but still exhibit phenomena that are not fully understood. One of these phenomena is the segregation of particles that constitute a dense granular flow, where particles separate due to differences in properties such as size and density. This work approaches segregation from a new perspective by developing a force model for a single particle based on the assumption that there are two opposing drag forces acting on the particle. Validation of the model in mono-disperse flows reveals a family of velocity profiles, including a Bagnold profile. Further validation is done by comparing with known particle scaling laws in literature. The force model shows that the mechanism of segregation is always present in dense granular flows and that mono-disperse flows are a unique case where it is not distinctly visible. The insights in this paper prove a fundamental new way of thinking about dense granular flows and its individual particles, opening the field for new research directions.
\end{abstract}

\maketitle
\section{Introduction}
% Why segregation in dense granular flows matter
Particle segregation in dense granular flows has been a research topic for more than four decades~\cite{thornton2026modeling} due to its importance in industry and natural hazards. Understanding segregation enables efficient mixing operations on industrial scale, and improve predictions of natural hazards such as avalanches and landslides. Predictions on this scale require continuum simulations methods such as transport equations or mixture theory~\cite{umbanhowar2019modeling,gray2005theory}. These methods require closure relations describing the interaction between the different granular flow constituents~\cite{barker2021coupling}. While many different closure relations have been proposed~\cite{tunuguntla2017comparing,bancroft2021drag,trewhela2021large}, it is still not fully understood why particles segregate.

% Intruder research
In an attempt to advance the fundamental understanding of segregation, an emerging research field focusses on the micro-mechanical modelling of a single intruder in dense granular flows. This intruder is varied in size, density or other particle properties with respect to the bulk particles. The objective is to find long-time averaged forces on the intruder, thereby revealing the underlying physics. A major breakthrough in this research avenue was the particle-on-a-spring method~\cite{guillard2016scaling} which enables measurements of the forces at specific points in the flow without the intruder going for a walk. This research area has resulted in the proposal of forces based on the local flow gradients such as the buoyancy, drag, lift, and segregation forces. Models for these forces are often inspired by forces observed in classical Newtonian fluids. 

%Buoyancy forces
The buoyancy force in Newtonian fluids is governed by Archimedes' principle, being equal to the displaced mass of the fluid, scaling with the pressure gradient of the fluid and acting in the opposite direction of the gravity. This force appears to be more complicated in dense granular flows due to the macroscopic size of the flow constituents. The pressure gradient in dense granular flows scales with the flow density, while the gravity force scales with the particle density. This difference causes a difference in magnitude between the Archimedes buoyancy force and the gravity force. As result, different models for the buoyancy force have been proposed to match this gap~\cite{lantman2021granular,yennemadi2023drag,jing2020rising}.

%drag forces
In viscous Newtonian fluids, particles experience Stokes' law, a drag force scaling with viscosity and a slip velocity (deviation of the intruder velocity w.r.t. the undisturbed flow). This type of force has been observed for intruders in dense granular flows in many different numerical simulations~\cite{yennemadi2023drag, tripathi2011numerical, he2025lift}, although the drag coefficient is different in  dense granular flows compared to classical fluids.

%Lift forces
Lift forces on intruders in fluids often are considered to be the force orthogonal to the flow direction, excluding the buoyancy force. In Ref.~\cite{PhysRevFluids.3.074303} a Saffman-like lift force was proposed. It was shown that the intruder experiences a lift force that is proportional to an observed horizontal slip velocity as the size ratio of the intruder increases. The relation between the lift force and slip velocity has been further investigated in a wider variety of flows~\cite{he2025lift,yennemadi2023drag}.

%segregation forces
The segregation force on an intruder is defined as the upward force acting on an intruder which includes any buoyancy and lift forces. Two generalised models for the segregation force have been proposed, one based on pressure and shear stress gradients~\cite{guillard2016scaling} and one based on pressure and shear rate gradients~\cite{jing2021unified}, both offering a different explanation for the origin of segregation.

Currently no single theory is able to link together all the different observed scaling laws for the forces on the intruder. Most likely because most proposed force models are either observed from exhaustive parameter exploration in simulations or ad-hoc model assumptions such as a Saffman lift force~\cite{PhysRevFluids.3.074303}. In this work a different approach is taken, first a micro-mechanical model is derived and then the emerging scaling laws are compared to observations from simulations.

This paper is organised as follows. In section~\ref{sec:derivation} a force model for an intruder is derived in which takes into account the local surroundings of an intruder particle, with specific focus on the drag force in section~\ref{sec:derivation_drag_force}. In section~\ref{sec:mododisperse_flows} the force model is first verified by considering mono-disperse flows. Lift forces in linear flows are considered in section~\ref{sec:lift_forces}, buoyancy forces in section~\ref{sec:buoyancy_forces} and segregation forces in section~\ref{sec:intruder_mechanics}.  Finally a discussion on the limitations of the model is given in section~\ref{sec:discussion} and in section~\ref{sec:conclusions} conclusions are drawn and an outlook is given.

\section{Intruder Force Model}
\label{sec:derivation}
In this work an intruder is considered to be a single particle that is submerged in a mono-disperse granular flow. The properties of the intruder may be equal to a bulk flow particle or uniquely different (e.g. different radius and density). The intruder may flow freely or is fixed with an external force (e.g. a spring force). In real-world applications an intruder will follow  Newton's second law in the form of
\begin{equation}
\bm{F}_m(\bm{x}_I,\dot{\bm{x}}_I) + \bm{F}_f = \rho_I V_I \ddot{\bm{x}_I},
\end{equation}
where $F_m$ are the long time-averaged mean forces on the intruder, which depend on the current position $\bm{x}_I$ and velocity $\dot{\bm{x}}_I$ of the intruder. The force $F_f$ is a stochastic contribution that takes fluctuations into account. This idea is similar to the Reynolds decomposition used in the Reynolds-Averaged Navier-Stokes equations, which decomposes the velocity into a mean and a fluctuating part. The right-hand side contains the intruder density $\rho_I$, volume $V_I$ and acceleration $\ddot{\bm{x}}_I$. The focus in this work is on the long time-averaged behaviour of the intruder such that the stochastic component can be ignored and the bulk particles around the intruder act like a continuum. The benefit of such a force model is that it can describe the mean behaviour of a single particle based on the average flow state around that particle. One application is that it can predict segregation velocities using a terminal velocity approach~\cite{Duan_Jing_Umbanhowar_Ottino_Lueptow_2025}.
 
In this section a model for $\bm{F}_m$ is developed. First an overview is given on the micro-mechanical environment around a single bulk particle. Based on this analysis a drag force is derived for an intruder with unspecified properties and finally a complete force model is proposed.

\subsection{Flow around a bulk particle}
\label{sec:flow_recap}
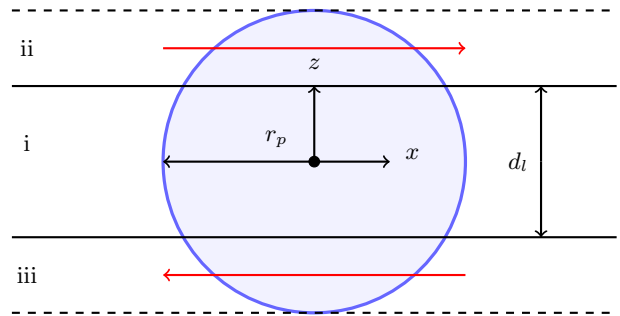
\begin{figure}
\begin{center}
\begin{tikzpicture}
\filldraw[color=blue!60, fill=blue!5, very thick](0,0) circle (2.0);
\filldraw[black] (0,0) circle (2pt);
\draw[black, thick] (-4,1) -- (4,1);
\draw[black, thick] (-4,-1) -- (4,-1);
\draw[black, dashed, thick] (-4,2) -- (4,2);
\draw[black, dashed, thick] (-4,-2) -- (4,-2);
\draw[red, <-, thick] (-2,-1.5) -- (2,-1.5);
\draw[red, ->, thick] (-2,1.5) -- (2,1.5);
\draw[black, ->, thick] (3,0) -- (3,-1);
\draw[black, ->, thick] (0,0) -- (1,0);
\draw[black, ->, thick] (3,0) -- (3,1);
\draw[black, ->, thick] (0,0) -- (0,1);
\node at (1.3,0.1) [black] {$x$};
\node at (0.0,1.3) [black] {$z$};
\node at (2.7,0.0) [black] {$d_l$};
\draw[black, ->, thick] (0,0) -- (-2,0);
\node at (-0.5,0.3) [black] {$r_p$};
\node at (-3.8,1.5) [black] {ii};
\node at (-3.8,0.25) [black] {i};
\node at (-3.8,-1.5) [black] {iii};
\end{tikzpicture}
\caption{\label{fig:layer_mechanics} A schematic long-time averaged view of a bulk particle with radius $r_p$ inside a layer moving in a dense granular flow with positive shear rate. The red arrows indicate the relative direction of the flow of the top and bottom layer. The thickness of the particle layer is $d_l$. Three distinctive regions are identified, i) top ion, ii) middle region and iii) bottom region. }
\end{center}
\end{figure}

Consider a simple granular flow which flows in the $x$-direction and only has gradients in the $z$-direction and a positive shear rate $\dot{\gamma}$. It has been shown for these flows that particles on average arrange themselves in flowing layers~\cite{forterre2008flows}. The average layer thickness $d_l$ is slightly smaller than twice the radius of a bulk particle $r_p$~\cite{weinhart2013coarse,thesisMarnix}, so there is some interaction between the layers. If a single bulk particle is considered inside such layer, its surrounding can be split into three regions, see Fig.~\ref{fig:layer_mechanics}. A middle layer i) where the particle mostly has interactions with horizontally neighbouring particles with the same velocity. A top layer ii) where the bulk particles travel faster than the middle layer and iii) a layer, where the particles travel slower than the middle layer.

It is reasonable to imagine that the interaction of a particle with a neighbouring layer acts on average like a drag force. With increasing velocity difference between the top layer and the middle layer a higher force on the intruder is to be expected. However, since there are gradients in the granular flow, the drag force in region ii) might be different from the drag force in region iii). Therefore, in this work two drag forces are proposed that act on the bulk particle. Naturally, these drag forces extend in some form to an intruder with properties that differ from the bulk particles. Interactions between the intruder and particles from region i) are assumed to have no net effect on the force balance. In the following section these drag forces are postulated for such a generic intruder.

\subsection{Drag force on an intruder}
\label{sec:derivation_drag_force}
Consider the numerical Stokes experiment from~\cite{tripathi2011numerical} where a Stokesian drag law was discovered for a free flowing intruder that had higher density, but similar size to the surrounding bulk particles. It traversed the flow faster than the surrounding particles and a drag law was found in the form of
\begin{equation}
\label{eq:basic_draw}
\bm{F} = -c \eta \bm{\lambda} R.
\end{equation}
Here $c$ is a drag coefficient, $\eta$ is the viscosity of the undisturbed flow (i.e., no intruder present), $R$ is the radius of the intruder and $\bm{\lambda}$ is the slip velocity of the intruder, 
\begin{equation}
\bm{\lambda} = \dot{\bm{x}}_I - \bm{v}(\bm{x}_I).
\end{equation}
Here $\bm{v}(\bm{x}_I)$ is the reference velocity of the undisturbed flow evaluated at the position of the intruder.
A positive slip velocity is when the intruder moves faster than the reference velocity. The form of this drag law will serve as base for the drag forces on the intruder in the top and bottom layer.
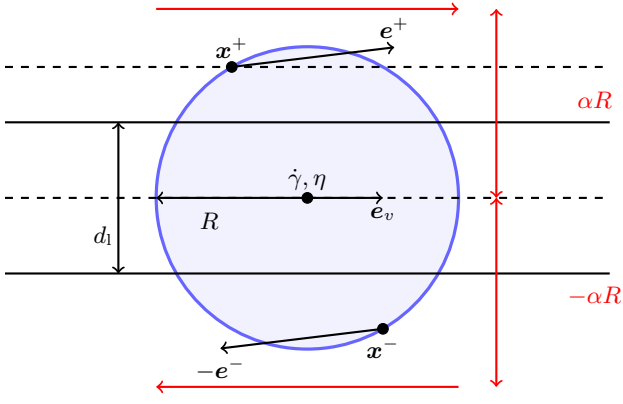
\begin{figure}
\begin{center}
\begin{tikzpicture}
\filldraw[color=blue!60, fill=blue!5, very thick](0,0) circle (2.0);
\filldraw[black] (0,0) circle (2pt) node[anchor=south]{$\dot{\gamma}, \eta$};
\draw[black, thick] (-4,1) -- (4,1);
\draw[black, thick] (-4,-1) -- (4,-1);
\draw[black, thick, dashed] (-4, 0) -- (4,0);
\draw[black, thick, dashed] (-4, {2.0 * sin(120)}) -- (4,{2.0 * sin(120)});
\draw[->, thick] (0,0) -- (1,0) node[anchor=north]{$\bm{e}_v$};
\filldraw[black] ({2.0 * cos(120)}, {2.0 * sin(120)}) circle (2pt) node[anchor=south]{$\bm{x}^+$};
\filldraw[black] ({2.0 * cos(300)}, {2.0 * sin(300)}) circle (2pt) node[anchor=north]{$\bm{x}^-$};
\draw[<->, black, thick] (-2.5, 1) -- (-2.5,-1);
\node at (-2.7,-0.5) {$d_{\textrm{l}}$};
\draw[->, black, thick] (0, 0) -- (-2,0);
\node at (-1.3,-0.3) [black] {$R$};
\draw[<->, red, thick] (2.5, 0) -- (2.5, 2.5);
\node at (3.8,1.3) [red] {$\alpha R$};
\draw[<->, red, thick] (2.5, 0) -- (2.5,-2.5);
\node at (3.8,-1.3) [red] {$-\alpha R$};
\draw[->, black, thick] ({2.0 * cos(120)}, {2.0 * sin(120)}) -- ({2.3 * cos(60)}, {2.3 * sin(60)}) node[anchor=south]{$\bm{e}^{+}$};
\draw[->, black, thick] ({2.0 * cos(300)}, {2.0 * sin(300)}) -- ({2.3 * cos(240)}, {2.3 * sin(240)}) node[anchor=north]{$-\bm{e}^{-}$};
% velocity arrows
\draw[red, <-, thick] (-2,-2.5) -- (2,-2.5);
\draw[red, ->, thick] (-2,2.5) -- (2,2.5);
\end{tikzpicture}
\caption{\label{fig:schematic} A schematic view of a generic intruder with radius $R$ that moves in the direction of $\bm{e}_v$ through a layer with size $d_l$. The reference flow properties $\eta$ and $\dot{\gamma}$ are evaluated at the centre of the intruder. The black lines indicate the average layer with a thickness of $d_l$. Two drag forces are applied on the intruder at a position $\bm{x}$ and have a unit direction of $\bm{e}$. Top and bottom region are denoted by superscript $+$ and $-$.  Furthermore, the distance from the intruder centre to a representable velocity is indicated by $\alpha R$. The direction of $\bm{e}$ is an approximation of what might be expected for a bulk particle.}
\end{center}
\end{figure}

Before defining the drag forces on the intruder it is important to note that a classical drag force acts by definition on the whole intruder and in the opposite direction of the slip velocity. In the situation sketched in Fig.~\ref{fig:layer_mechanics} this is not the case as the drag force at the top layer only acts on the part of the intruder which is exposed to the top layer and not the whole intruder. Therefore, the drag force caused by the \textit{horizontal} slip velocity may contain components in the \textit{vertical} direction. In Fig.~\ref{fig:schematic} the normalised drag forces caused by the horizontal slip velocity on the intruder are schematically depicted. The unit flow direction is $\bm{e}_v$. The drag forces are applied at application points $\bm{x}^{+}$ and $\bm{x}^{-}$ and the unit direction of the force with positive slip velocity is given by the vector $\bm{e}^+$ and $\bm{e}^-$. The true direction is governed by the sign of the local slip velocity. The exact application points $\bm{x}^+$ and $\bm{x}^-$ are not relevant for the force model as they can be linearly transformed to the centre of mass of the intruder. 
It is crucial to note that no assumptions are made on these positions and directions as they may change significantly depending on the intruder properties and local flow configuration. However, it is likely that the application points will be somewhere in the top and bottom layer. The directions for a bulk particle are approximately in the direction of the relative layer velocity, as depicted in Fig.~\ref{fig:schematic} and since the drag forces are only acting locally on the intruder, they are allowed to have vertical component.
The drag forces may induce rotation on the intruder, however this is not explicitly considered here for simplicity. A recent study shows that for unforced rotating intruders this is a fair assumption~\cite{he2025lift}. 

First consider the drag force at the top-side caused by the horizontal velocity of that region. The drag force from Eq.~\eqref{eq:basic_draw} takes the form of
\begin{equation}
\label{eq:top_drag}
\bm{F}_{d,\lambda^{+}_x} = -c \eta^{+} \lambda^{+}_x R \bm{e}^{+}.
\end{equation}
Here $\eta^+$ and $\lambda^{+}_x$ are the viscosity and horizontal slip velocity for the top region. The slip velocity can be approximated by means of a first order Taylor expansion of the velocity about the centre of the intruder,
\begin{equation}
\label{eq:drag_force_top_slip_velocity}
\lambda^+_x = \dot{x}_{I,x} - v_x \left( \bm{x}_{I} + 
\begin{bmatrix} 
0 \\
\alpha R
\end{bmatrix}
\right) \approx \lambda_x - \dot{\gamma} \alpha R.
\end{equation}
The term $\alpha R$ indicates the height from the centre of the intruder at which the velocity has to be evaluated. The exact formulation of $\alpha$ is not known a-priori, however it depends likely on the size-ratio $S = R / r_p$. 

The viscosity can also be approximated with a first order Taylor expansion of the viscosity,
\begin{equation}
\eta^{+} = \eta + \dz{\eta} \alpha R,
\end{equation}
Expanding all terms in Eq.~\eqref{eq:top_drag} leads to
\begin{equation}
\bm{F}_{d,\lambda^{+}_x} = -c\left( 
\eta \lambda_x R
- \eta \dot{\gamma} \alpha R^2
+ \lambda_x \dz{\eta} \alpha R^2
- \dot{\gamma}\dz{\eta} {\alpha}^2 R^3
  \right) \bm{e}^{+}
\end{equation}

A similar approach can be taken for the bottom region,
\begin{equation}
\bm{F}_{d,\lambda^{-}_x} = -c \eta^{-} \lambda^{-}_x R \bm{e}^{-},
\end{equation}
but now the Taylor expansions are done in the opposite direction, $ -\alpha R$ from the centre of the intruder. The resulting drag force for the bottom region therefore is
\begin{equation}
\bm{F}_{d,\lambda^{-}_x} = -c \left( 
\eta \lambda _x R
+ \eta \dot{\gamma} \alpha R^2
- \lambda_x \dz{\eta} \alpha R^2
- \dot{\gamma}\dz{\eta} {\alpha}^2 R^3
  \right)  \bm{e}^{-}.
\end{equation}
The total drag force caused by the horizontal slip velocity is $\bm{F}_{d,\lambda_x} = \bm{F}_{d,\lambda^{+}_x}  + \bm{F}_{d,\lambda^{-}_x}$,
\begin{align}
\label{eq:horizontal_drag_force}
\bm{F}_{d,\lambda_x} = -c \left( 
\left(\eta \lambda_x R - \dot{\gamma}\dz{\eta} \alpha^2 R^3 \right) \left[\bm{e}^{+} + \bm{e}^{-} \right] \right.\nonumber \\
\left. + \left(\lambda_x \dz{\eta} - \tau \right) \alpha R^2 \left[\bm{e}^{+}  - \bm{e}^{-} \right]
  \right)
\end{align}
where $\tau = \eta \dot{\gamma}$ is the shear stress. 

To get a better understanding of this drag force, consider an intruder identical to a bulk particle. The change in angle between $\bm{e}^+$ and $\bm{e}^-$ can only change slightly over the length of the intruder as they will remain mostly in the direction of the flow. Therefore, a simplified approximation may be obtained by assuming $\bm{e}^- = \bm{e}^+$, leading to
\begin{equation}
\tilde{\bm{F}}_{d,x} = -2 c \left(
\eta \lambda_x R  - \alpha^2 \dot{\gamma}\dz{\eta}  R^3 
\right) \bm{e}^+.
\end{equation}
Note that the first term is the standard Stokesian drag which was the base assumption of the drag force. The second term is new and stems from the difference in magnitude of the drag forces in the top and bottom region. The term describes the difference between velocity and viscosity in the top and bottom layer, hence this term scales with both gradients $\dz{\eta}$ and $\dot{\gamma}$. Observe that the first term scales with the intruder radius, while the second term scales with the cubed radius which is unexpected from a drag force.

The drag force related to a \textit{vertical} slip velocity $\lambda_z$ is taken to be simple Stokesian drag force~\cite{Jing_Ottino_Umbanhowar_Lueptow_2022},
\begin{equation}
\bm{F}_{d,\lambda_z} = -c \eta \lambda_z R \bm{e}_z,
\end{equation}
as there are no gradients present in the $x$-direction of the granular flow. It is assumed that $c$ in the $x$-direction is the same as the $z$-direction. In the next section, a full force model for a generic intruder is proposed.

\subsection{Force Model}
\label{sec:force_model}
A full description of a generic intruder force model contains several (optional) force contributions, an external force $\bm{F}_e$ (e.g., a spring force), the gravity force $\bm{F}_g$, buoyancy force $\bm{F}_b$ and the drag force $\bm{F}_d$,
\begin{equation}
\bm{F}_e + \bm{F}_g + \bm{F}_b +  \bm{F}_{d,\lambda_x} + \bm{F}_{d,\lambda_z} = \rho_I V_I \ddot{\bm{x}_I}.
\end{equation}
There is no explicit lift force defined in this force balance as the vertical contribution of $\bm{F}_{d,\lambda_x}$ can be interpreted as a lift force.

The gravity force is defined as
\begin{equation}
\bm{F}_g = \rho_p \bm{g} V_I,
\end{equation}
where $\rho_p$ is the density of the intruder. For the buoyancy force, a different approach is taken to other approaches in literature~\cite{jing2020rising,yennemadi2023drag,lantman2021granular}. Here the buoyancy force is simply taken as the classical Archimedes buoyancy force by integrating the macroscopic stress-tensor $\bm{\sigma}$ of the granular flow over the surface of the particle,
\begin{equation}
\bm{F}_{b} = \int \bm{\sigma} \bm{n} dA.
\end{equation}
For flows considered in this work, the buoyancy force is
\begin{equation}
\bm{F}_{b} = \int \bm{\sigma} \bm{n} dA =  
\begin{bmatrix}
\dz{\tau} \\
- \dz{p}
\end{bmatrix}
V_I
\end{equation}
where $\dz{\tau}$ is the gradient in shear stress and $\dz{p}$ is the hydrostatic pressure gradient and follow from the macroscopic force balance with gravity,
\begin{equation}
\label{eq:dztau}
\dz{\tau} = - \phi \rho_f g_x,
\end{equation}
and
\begin{equation}
\label{eq:dpdz}
\dz{p} = \phi \rho_f g_z,
\end{equation}
with $\rho_f$ as the density of the flow particles and $\phi$ the solid volume fraction. The buoyancy force therefore becomes
\begin{equation}
\bm{F}_b = - \phi \rho_f \bm{g} V_I.
\end{equation}
Note that this force is smaller in magnitude than the gravity force. The gravity and buoyancy force can be combined,
\begin{equation}
\bm{F}_g + \bm{F}_b = (D - \phi) \rho_f \bm{g} V_I,
\end{equation}
where $D = \rho_p / \rho_f$ is the density ratio. The final force model then becomes

\begin{align}
\label{eq:total_force_model}
 \bm{F}_e + (D - \phi) \rho_f \bm{g} V_I
 &-c \biggl( 
\left(\eta \lambda_x R - \dot{\gamma}\dz{\eta} \alpha^2 R^3 \right) \left[\bm{e}^{+} + \bm{e}^{-} \right] \nonumber \\
&+\left(\lambda_x \dz{\eta} - \tau \right) \alpha R^2 \left[\bm{e}^{+}  - \bm{e}^{-} \right] \nonumber \\
  &+ \eta \lambda_z R \bm{e}_z \biggl)
  = \rho_I V_I \ddot{\bm{x}_I}.
\end{align}
This model is profoundly different to other proposed models when considering a bulk particle (i.e., $D = 1$ and $S=1$), as it is common to assume that the gravity and buoyancy force balance for mono-disperse particles by proposing a modified buoyancy force. Instead, this model proposes that there is always a drag force present which matches the difference between the gravity and the Archimedes buoyancy force. It then follows that the drag force must scale with both the pressure and the shear stress gradients. As a result, the model puts constraints on the behaviour of the granular flow as these stress gradients are now directly linked to shear rate and viscosity effects. These constraints on mono-disperse flows are explored in section~\ref{sec:mododisperse_flows}.

\section{Mono-disperse flows}
\label{sec:mododisperse_flows}
\subsection{Flow equation}
When the flow is mono-disperse, the force model should be the same for each particle inside the flow. Therefore, it is not possible to have a slip velocity and acceleration. Furthermore, there are no external forces present and the density ratio is one, reducing Eq.~\eqref{eq:total_force_model} to
\begin{align}
\label{eq:base_equation}
(1 - \phi) \rho_f \bm{g} V_I  + c \biggl(
\dot{\gamma}\dz{\eta} \alpha^2 R^3 \left[\bm{e}^{+} + \bm{e}^{-} \right] \nonumber \\
+ \tau \alpha R^2 \left[\bm{e}^{+}  - \bm{e}^{-} \right] \biggl) = \bm{0}.
\end{align}
The force balance in the orthogonal direction of $\bm{g}$ is obtained by taking the 2D cross product (i.e., $\bm{a} \times \bm{b} = a_x b_y - a_y b_x$) with $\bm{e}_g$, cancelling out the gravity term,
\begin{equation}
c \left(
\dot{\gamma}\dz{\eta} \alpha^2 R^3 \left[\bm{e}^{+} + \bm{e}^{-} \right]
+\tau \alpha R^2 \left[\bm{e}^{+}  - \bm{e}^{-} \right]
  \right) \times \bm{e}_g = 0.
\end{equation}
Dropping the factor $c$, this equation can be rewritten as
\begin{equation}
\label{eq:orthogonal_balance}
\tau = \alpha \dot{\gamma}\dz{\eta} R \psi,
\end{equation}
with
\begin{equation}
\label{eq:psi}
\psi = -\frac{\left[\bm{e}^{+} + \bm{e}^{-} \right] \times \bm{e}_g}
{\left[\bm{e}^{+}  - \bm{e}^{-} \right] \times \bm{e}_g}.
\end{equation}
Observe that Eq.~\eqref{eq:orthogonal_balance} can be used to eliminate $\tau$ in Eq.~\eqref{eq:base_equation},
\begin{equation}
\label{eq:flow_equation_vector_form}
(1 - \phi) \rho_f \bm{g} V_I + \frac{3}{4 \pi} c \alpha^2 \dot{\gamma}\dz{\eta} \bm{k} V_I = \bm{0},
\end{equation}
where $R^3$ has been rewritten to $V_I$ and the non-unit vector
\begin{equation}
\label{eq:drag_vector}
\bm{k} = \left[\bm{e}^{+} + \bm{e}^{-} \right] + \psi \left[\bm{e}^{+}  - \bm{e}^{-} \right].
\end{equation}
Taking the dot product with $\bm{k}$ reduces the equation to
\begin{equation}
(1 - \phi) \rho_f |\bm{g}||\bm{k}| \cos(\theta_{gk}) V_I + \frac{3}{4 \pi } c \alpha^2 \dot{\gamma}\dz{\eta} V_I = 0,
\end{equation}
where $\theta_{gk}$ is the angle between $\bm{e}_g$ and $\bm{k}$. After division with $\alpha^2 V_I$ the final form of the flow equation is obtained,
\begin{equation}
\label{eq:flow_equation}
\dot{\gamma}\dz{\eta} = (\phi - 1) \rho_f |\bm{g}| f_c,
\end{equation}
with
\begin{equation}
f_c = \frac{4 \pi}{3} \frac{|\bm{k}| \cos(\theta_{gd})}{c \alpha^2}.
\end{equation}
There are a few quantities in the right-hand-side of Eq.~\eqref{eq:flow_equation} that may vary based on the local flow properties. The solid fraction $\phi$ is a linear function of the Inertial number $I$~\cite{gdr2004dense},
\begin{equation}
I = d_p\dot{\gamma} \sqrt{\rho_f / p}
\end{equation}
and perhaps $\alpha$, and $\theta_{gk}$ change as function of $I$. However, all these quantities have a very narrow range of validity. Layers can't grow thicker than the particle diameter, and the drag forces are only possible in approximately the layer velocity direction. Therefore, as a first approximation, it is assumed that the right-hand-side of the flow equation is constant. Observe that the force model has now been transformed to an ordinary differential equation. In the next section solutions of this simplified flow equation are presented.

\subsection{Velocity profiles}
\label{sec:velocity_profiles}
\begin{figure}
\begin{center}
\includegraphics[width=0.99\linewidth]{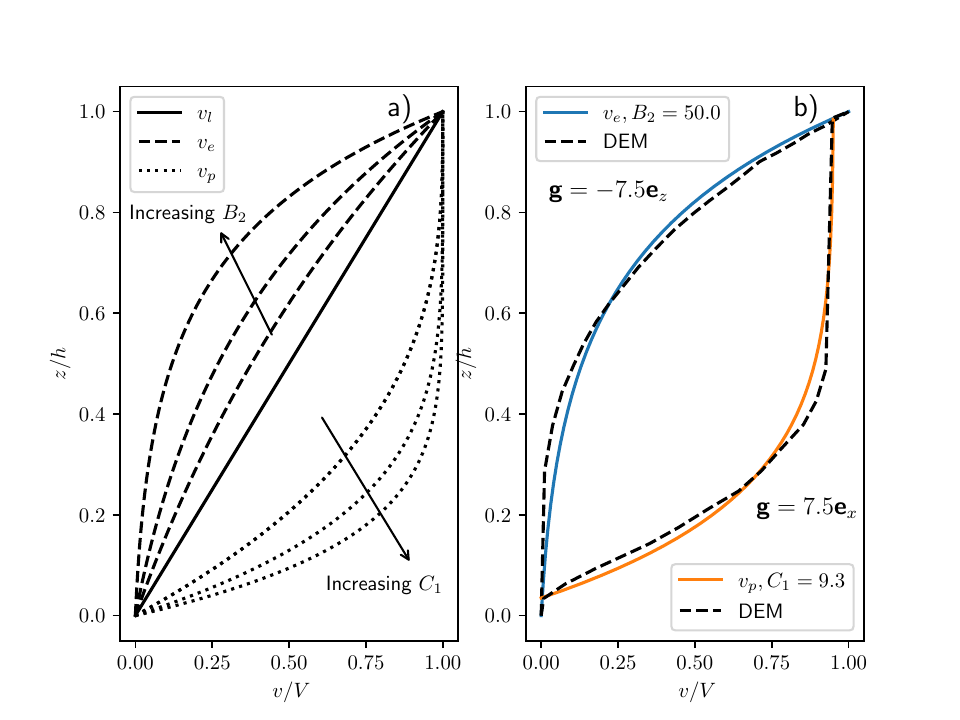}
\caption{\label{fig:family_of_curves} Velocity profiles as solution to Eq.~\eqref{eq:flow_equation}. In panel a) exponential, power and linear velocity profiles are shown. Boundary conditions taken are $v(0) = 0$, $v(h) = 1$, $h=1$ and $\bar{\tau} = 1$. Three different solutions are shown for Eq.~\eqref{eq:velocity_exponential_equations} with $B_2 = \{1,2,4 \}$ and Eq.~\eqref{eq:velocity_power_equation} with  $C_1 = \{3,5,7\}$. Panel b) shows a comparison of the velocity profiles with DEM results from Ref.~\cite{guillard2016scaling} for two confined flow cases with boundary conditions $v_0 = 0$ and $v_h = 2.5 \textrm{ m/s}$. }
\end{center}
\end{figure}

From the flow equation, Eq.~\eqref{eq:flow_equation}, it is possible to derive three different velocity profiles when the right-hand-side is constant, depending on $\bm{g}$. Solutions to this equation are derived in Appendix~\ref{sec::appendix_a}. The first solution is the trivial linear solution, when $|\bm{g}| = 0$,
\begin{equation}
v_{\textrm{l}}(z) = v_0 + \frac{v_h - v_0}{h}z,
\end{equation}
here $h$ is the height of the flow, $v(0) = v_0$ and $v(h) = v_h$. This solution does not depend on any micro-mechanical coefficients since it follows from pure symmetry between the competing drag forces. The second solution is an exponential solution for $g_x = 0, g_z > 0$,
\begin{equation}
\label{eq:velocity_exponential_equations}
v_{e}(z) = v_0 +  (v_h - v_0) \frac{1 - e^{B_2 z}}{1 - e^{B_2 h}},
\end{equation}
with
\begin{equation}
B_2 = \frac{\rho_p|g_z| (\phi - 1) f_c}{\tau_0}.
\end{equation}
The final equation is a power equation for $g_x > 0$,
\begin{equation}
\label{eq:velocity_power_equation}
v_p(z) = v_0 + (v_h - v_0) \frac{(\bar{\tau} - z)^{C_1} - \bar{\tau}^{C_1}}{(\bar{\tau} - h)^{C_1} - \bar{\tau}^{C_1}}.
\end{equation}
with 
\begin{equation}
C_1 = \frac{|\bm{g}|}{g_x} (1 - 1 / \phi) f_c
\end{equation}
and
\begin{equation}
\bar{\tau} = \frac{\tau_0}{\phi \rho_f g_x}.
\end{equation}
The flow profiles of these solutions are shown in Fig.~\ref{fig:family_of_curves} a) for various coefficients. The higher the coefficients, the stronger the deviation from the linear profile. 

To verify that these solutions are physically correct they are compared to DEM simulations in two distinct scenario's, shown in Fig.~\ref{fig:family_of_curves}b). The first scenario only has gravity in the $z$-direction, which requires $v_e$  and one with only gravity in the $x$-direction which requires $v_p$. The velocity profiles are compared to discrete element simulation data from Ref.~\cite{guillard2016scaling}. In both scenario's there is good comparsion. The horizontal gravity case  shows boundary effects and therefore the height of the flow and $v_h$ are adjusted to match the boundary conditions.

For free surface flows the stress vanishes at the surface of the flow. This can be achieved by setting $\bar{\tau} = h$, which yields the velocity profile
\begin{equation}
v_{\textrm{bag}}(z) = v_0 + \frac{(v_h - v_0)}{h^{C_1}} (h^{C_1} - (h - z)^{C_1}).
\end{equation}
Observe that this equation is of similar structure as a Bagnold flow that is allowed to slip~\cite{Wang_Jing_Kwok_Sobral_Weinhart_Thornton_2025} when $C_1 = 3/2$ and
\begin{equation}
v_h = v_0 + \frac{2 I \sqrt{\phi g \cos \theta}}{3d} h^{3/2}.
\end{equation}

\subsection{Remarks}
The flow equation describes a general family of velocity profiles. The exact velocity profile depends on the micro-mechanical properties which is governed by the parameter $f_c$, resulting from the two competing drag forces. The comparison of the flow equation with DEM verifies that it has the capability to reproduce the velocity profile of different DEM cases with only a single fitting parameter. The curious observation here is that the mathematical structure of the velocity profiles can be derived without using the $\mu(I)$ rheology, but follows from the long time-averaged competing drag forces on a particle. This is a novel way to describe granular flows and might prove useful for further understanding the origin of $\mu(I)$-rheology.

While the flow equation generates the right structure, it still requires closure relations to fully resolve velocity profile. The angle $\theta_{gk}$ is not predicted by the current theory and relies on fitting the velocity profile to numerical simulations. In addition, for free-surface flows the flow equation $v_h$ is not known. The model contains the assumption that the flow quantities $\phi$, $\alpha$ and $\theta_{gk}$ remain constant. This flow model can be refined by more precise modelling of these quantities and more detailed study of Eq.~\eqref{eq:orthogonal_balance} and Eq.~\eqref{eq:psi}. However, that is beyond the scope of this work. The derived velocity profiles give give strong evidence that Eq.~\eqref{eq:total_force_model} is the correct force model.

\section{Lift and drag forces in linear flows}
\label{sec:lift_forces}
In the context of size segregation, lift forces $F_L$ are considered as the part of the segregation force that is not the buoyancy force. In Ref.~\cite{PhysRevFluids.3.074303} a first attempt was made to understand this lift force on an intruder in chute flow by defining a Voronoi-based buoyancy force model and a Saffman lift force. A slip velocity $\lambda_x$ was found in the flow direction and a scaling of $\lambda_x \propto F_L$ was observed. Even though the Saffman lift force could not be generalised to other systems~\cite{yennemadi2023drag}, the connection between the observed horizontal slip velocity and the lift force was an important finding that has been studied in more detail in Ref.~\cite{he2025lift} for linear flow profiles. The benefit of their approach is that there is no buoyancy force present, which allows to study the lift and drag forces explicitly without having to assume a decomposition. 

In the study a linear flow profile was imposed while the intruder was fixed. By varying the offset of the linear profile a slip velocity could be applied. The general force balance proposed in Ref.~\cite{he2025lift} in $z$-direction is
\begin{equation}
\label{eq:linear_system}
\begin{bmatrix}
F_p \\
F_s
\end{bmatrix}
+
\begin{bmatrix}
\tilde{F}_d \\
F_L 
\end{bmatrix}
= \bm{0}
\end{equation}
where $F_p$ and $F_s$ are the measured forces in the, respectively, $x$- and $z$-direction that keep the intruder in its place, $\tilde{F}_d$ is the resulting drag force in horizontal direction and $F_L$ is the lift force. The tilde for the drag force is introduced to distinguish this drag force definition from the drag forces in the model. The next section applies the force model to this system

\subsection{Force Model}
When applying the force model, Eq.~\eqref{eq:total_force_model}, to the linear system, the following observations are made.
First, the external forces imposed on the intruder are $F_p$ and $F_s$. Secondly, due to a linear flow profile there are no viscosity gradients present and finally the force in the $z$-direction holds the intruder in place such that $\lambda_z$ and $\ddot{\bm{x}}$ are zero. Therefore, the reduced force model becomes
\begin{equation}
\label{eq:linear_force_model_a}
\begin{bmatrix}
F_p \\
F_s 
\end{bmatrix}
- c \eta R \lambda_x \left[\bm{e}^+ + \bm{e}^- \right] 
+ c\tau R^2 \alpha \left[\bm{e}^{+}  - \bm{e}^{-} \right] = \bm{0}.
\end{equation}
Comparing Eq.~\eqref{eq:linear_force_model_a} with Eq.~\eqref{eq:linear_system} gives the definition of the forces
\begin{equation}
\label{eq:lift_drag_base_equation}
\begin{bmatrix}
\tilde{F}_d \\
F_L
\end{bmatrix}
= -c \eta R \lambda_x \left[\bm{e}^+ + \bm{e}^- \right] 
+ c\tau R^2 \alpha \left[\bm{e}^{+}  - \bm{e}^{-} \right].
\end{equation}
These equations can be simplified further due to a symmetry condition in the linear system. Consider the lift/drag ratio $R_{LD} = -F_L / F_d$.
The symmetry condition is $R_{LD}(\lambda) = R_{LD}(-\lambda)$, the ratio between the lift and drag for a positive slip velocity should be equal to the lift drag ratio for the same slip velocity in the other direction which was verified in Ref.~\cite{he2025lift}. This symmetry condition can only be satisfied if the second term of Eq.~\eqref{eq:lift_drag_base_equation} is zero. This is only possible when $\bm{e}^+ = \bm{e}^-$.
Hence Eq.~\eqref{eq:lift_drag_base_equation} can be reduced to
\begin{equation}
\label{eq:lift_drag_improved_equation}
\begin{bmatrix}
\tilde{F}_d \\
F_L
\end{bmatrix}
= -2c \eta R \lambda_x \bm{e}^+.
\end{equation}
In the next section this equation is compared to measurements.

\subsection{Scaling}
\begin{figure}
\begin{center}
\includegraphics[width=0.99\linewidth]{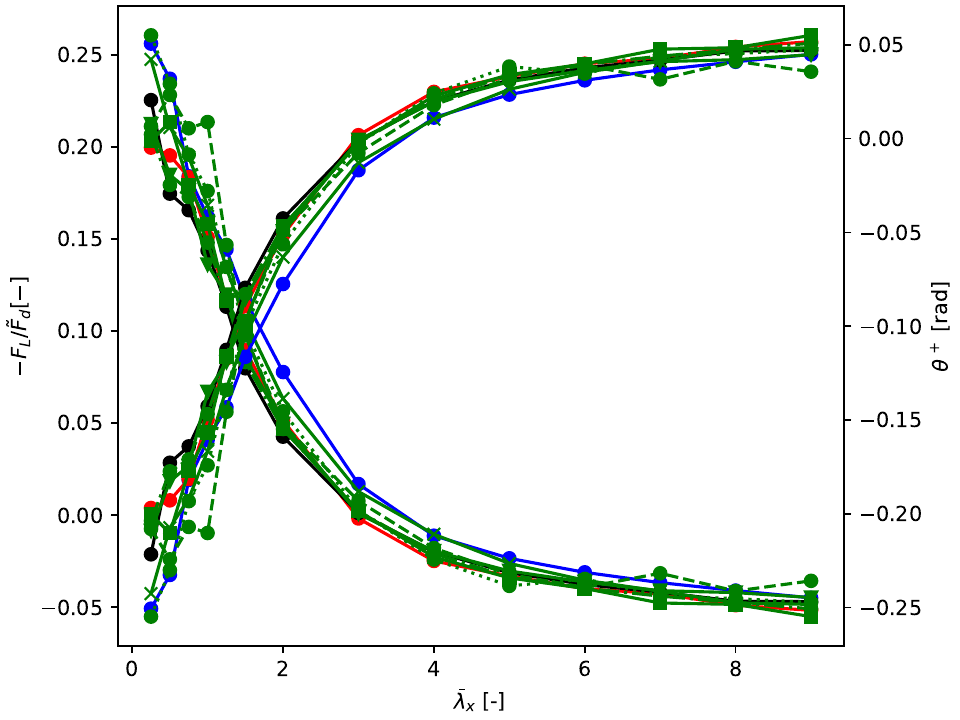}
\caption{\label{fig:angles} The left axis of this plot contains measured lift/drag ratio from~\cite{he2025lift} for various $p$, $\dot{\gamma}$ and $S$. Values for $p$ are \{diamond=250, square=500, cross=2000, dot=1000\}, values for $\dot{\gamma}$ are \{blue=2.5, red=5, black=10\} and values for $S$ are \{dashed=1, dotted=2, solid=3, dashdot=4\}. The right axis shows the angle of the drag force, $\theta^+$.}
\end{center}
\end{figure}

The scaling of various properties of the linear force model, Eq~\eqref{eq:lift_drag_improved_equation} will be investigated. One interesting aspect is that in a linear system it is possible to measure the angle $\theta^+$ of vector $\bm{e}^+$. Taking the ratio between the lift and drag component of Eq~\eqref{eq:lift_drag_improved_equation} yields
\begin{equation}
R_{LD} = -\tan \theta^+.
\end{equation}
The measured lift/drag ratios from Ref.~\cite{he2025lift} are shown in Fig.~\ref{fig:angles} together with the angle $\theta^+$ as function of the non-dimensional slip velocity $\bar{\lambda}$ defined in Ref.~\cite{he2025lift} as
\begin{equation}
\label{eq:non_dim_slip_velocity}
\bar{\lambda_x} = \frac{\lambda_x}{\dot{\gamma}R(1 + 1/S)}
\end{equation}
The lift/drag ratio starts out postive, leading to an upward force on the intruder. As the $\bar{\lambda}$ increases, the ratio crosses the zero point at roughly $\bar{\lambda}_x \approx 3.5$ implying that the lift force is zero. Further increase of $\bar{\lambda}$ shows that the particle experiences a downward force and it saturates at roughly -0.03. At zero slip velocity $\theta^+$ starts out at an angle of -0.25 radians and saturates at an angle of 0.05 radians. The clean, seamingly collapsed, curve of $R_{LD}$ suggests that $\theta^+$ is mostly a function of $\bar{\lambda}_x$.

\begin{figure}
\begin{center}
\includegraphics[width=0.99\linewidth]{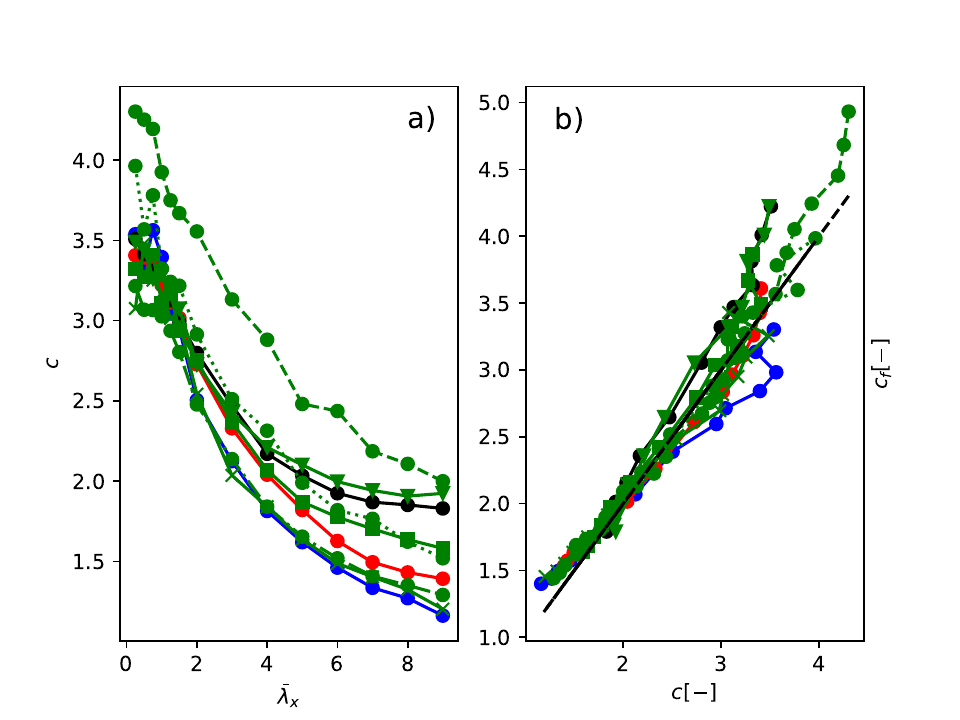}
\caption{\label{fig:c} In panel a) the coefficient $c$ of Eq.~\eqref{eq:lift_drag_improved_equation} is shown. The colours, markers and line styles are similar to Fig.~\ref{fig:angles}. In panel b) the emperical fit Eq.~\eqref{eq:emperical_fit_c} is plotted against the measured data. The black dashed line indicates a perfect match.}
\end{center}
\end{figure}

With the angle known it is now possible to investigate the scaling of $c$ in Eq.~\eqref{eq:lift_drag_improved_equation}. The viscosity in this equation is estimated using the $\mu(I)$-rheology with $\mu(I) = 0.36 + 0.55 / (0.73 / I + 1)$. In Fig.~\ref{fig:c}a) the coefficient is plotted as function of $\bar{\lambda}$. The value is approximately around 3 for zero slip velocity and reduces for higher values of $\bar{\lambda}$. There are three relevant non-dimensinoal parameters: $S$, $I$ and $\bar{\lambda}$. A fit of $c$ is obtained by investigating each dependency separate, leading to an empirical equation of
\begin{equation}
\label{eq:emperical_fit_c}
c_f = (1 + S^{-0.7})(1.35 + 7.0I)(0.6 + \exp^{-0.35 \bar{\lambda_x}}).
\end{equation}
In Fig~\ref{fig:c}b) a comparison between $c$ and $c_f$ is shown, which gives decent agreement. There is some spread at higher values of $c$ which corresponds to measurements at small slip velocity where $c$ fluctuates significantly. The size-ratio dependency in $c_f$ may be related to the reduction in surface contact density of the intruder~\cite{lantman2021granular}.

In Ref.~\cite{he2025lift} it was proposed that the lift force scales as $F_L(\bar{\lambda}_x) \propto pR^2(1 + S^{-1})^2$. It can be shown that this is indeed part of the complete scaling when the non-dimensional slip velocity, Eq.~\eqref{eq:non_dim_slip_velocity}, is substituted in Eq.~\eqref{eq:lift_drag_improved_equation},
\begin{equation}
F_L = -2 c \mu p R^2 (1 + 1/S) \bar{\lambda_x} \bm{e}^+.
\end{equation}
When taking the $S$ dependency from $c$ this yields a similar scaling of $p R^2 (1 + S^{-1})(1 + S^{-0.7})$.

\subsection{Interpretation}
\begin{figure}
\begin{center}
\includegraphics[width=0.99\linewidth]{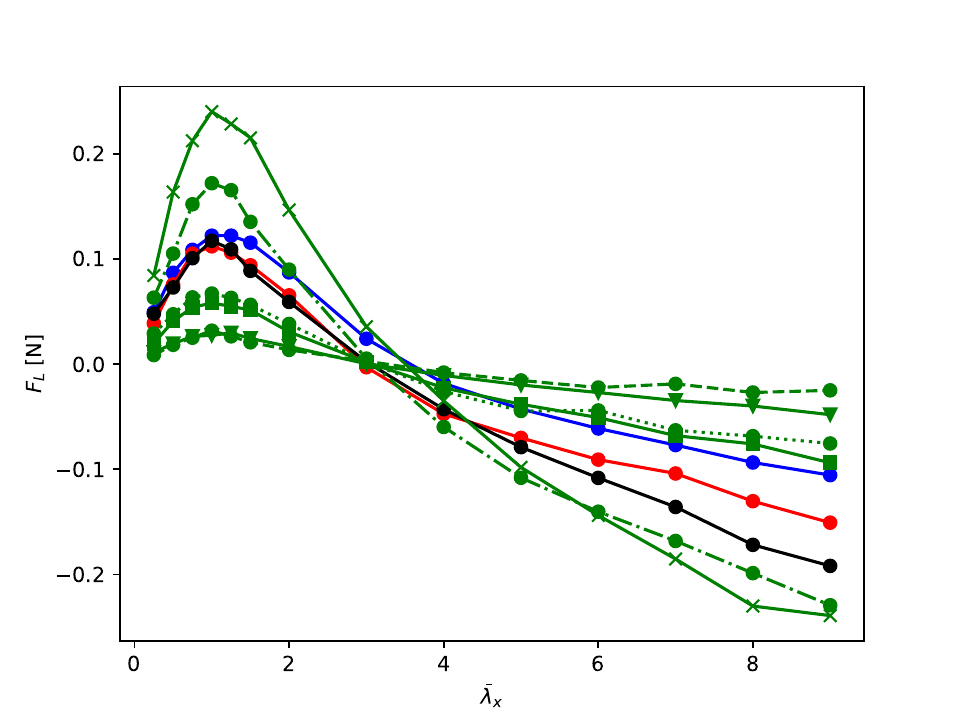}
\caption{\label{fig:lift_forces} The measured lift forces from Ref.~\cite{he2025lift} are shown. The colours, markers and line styles are similar to Fig.~\ref{fig:angles}.}
\end{center}
\end{figure}

\begin{figure*}
\begin{center}
\begin{tikzpicture}
% circles
\filldraw[color=blue!60, fill=blue!5, very thick](0,0) circle (1.5);
\filldraw[color=blue!60, fill=blue!5, very thick](4,0) circle (1.5);
\filldraw[color=blue!60, fill=blue!5, very thick](8,0) circle (1.5);
% circle centres
\filldraw[black] (0,0) circle (2pt);
\filldraw[black] (4,0) circle (2pt);
\filldraw[black] (8,0) circle (2pt);
% centreline and layer thickness
\draw[black, thick, dashed] (-2, 0) -- (10, 0);
\draw[black, thick] (-2, 0.8) -- (10, 0.8);
\draw[black, thick] (-2, -0.8) -- (10, -0.8);
% x_p-
\pgfmathsetmacro{\xam}{1.5 * cos(305)};
\pgfmathsetmacro{\yam}{1.5 * sin(305)};
\pgfmathsetmacro{\xbm}{1.5 * cos(305) + 4};
\pgfmathsetmacro{\ybm}{1.5 * sin(305)};
\pgfmathsetmacro{\xcm}{1.5 * cos(305) + 8};
\pgfmathsetmacro{\ycm}{1.5 * sin(305)};
\filldraw[black] (\xam, \yam) circle (2pt) node[anchor=north]{$\bm{x}^-$};
\filldraw[black] (\xbm, \ybm) circle (2pt) node[anchor=north]{$\bm{x}^-$};
\filldraw[black] (\xcm, \ycm) circle (2pt) node[anchor=north]{$\bm{x}^-$};
% x_p+ 
\pgfmathsetmacro{\xap}{1.5 * cos(125)};
\pgfmathsetmacro{\yap}{1.5 * sin(125)};
%\pgfmathsetmacro{\xbp}{1.5 * cos(125) + 4};
%\pgfmathsetmacro{\ybp}{1.5 * sin(125)};
\pgfmathsetmacro{\xcp}{1.5 * cos(125) + 8};
\pgfmathsetmacro{\ycp}{1.5 * sin(125)};
\filldraw[black] (\xap, \yap) circle (2pt) node[anchor=south]{$\bm{x}^+$};
%\filldraw[black] (\xbp, \ybp) circle (2pt) node[anchor=south]{$\bm{x}^+$};
\filldraw[black] (\xcm, \ycp) circle (2pt) node[anchor=west]{$\bm{x}^+$};
% e-
\pgfmathsetmacro{\xamm}{\xam + 1.0 * cos(165.68)};
\pgfmathsetmacro{\yamm}{\yam + 1.0 * sin(165.68)};
\pgfmathsetmacro{\xbmm}{\xbm + 1.0 * cos(171.41)};
\pgfmathsetmacro{\ybmm}{\ybm + 1.0 * sin(171.41)};
\pgfmathsetmacro{\xcmm}{\xcm + 1.0 * cos(174.28)};
\pgfmathsetmacro{\ycmm}{\ycm + 1.0 * sin(174.28)};
\draw[->, black, thick] (\xam, \yam) -- (\xamm, \yamm) node[anchor=east]{$-\bm{e}^{-}$};
\draw[->, black, thick] (\xbm, \ybm) -- (\xbmm, \ybmm) node[anchor=east]{$-\bm{e}^{-}$};
\draw[->, black, thick] (\xcm, \ycm) -- (\xcmm, \ycmm) node[anchor=east]{$-\bm{e}^{-}$};
% e+
\pgfmathsetmacro{\xapp}{\xap + 1.0 * cos(-14.32)};
\pgfmathsetmacro{\yapp}{\yap + 1.0 * sin(-14.32)};
%\pgfmathsetmacro{\xbpp}{\xbp + 1.0 * cos(350)};
%\pgfmathsetmacro{\ybpp}{\ybp + 1.0 * sin(350)};
\pgfmathsetmacro{\xcpp}{\xcp + 1.0 * cos(174.28)};
\pgfmathsetmacro{\ycpp}{\ycp + 1.0 * sin(174.28)};
\draw[->, black, thick] (\xap, \yap) -- (\xapp, \yapp) node[anchor=west]{$\bm{e}^{+}$};
%\draw[->, black, thick] (\xbp, \ybp) -- (\xbpp, \ybpp) node[anchor=south]{$\bm{e}^{+}$};
\draw[->, black, thick] (\xcm, \ycp) -- (\xcmm, \ycpp) node[anchor=east]{$-\bm{e}^{+}$};
% layer thickness
\draw[<->, black, thick] (-1.8, -0.8) -- (-1.8, 0.8);
\node at (-2.0, 0.2){$d_l$};
% Relative velocity
\draw[->, red, thick] (-0.5, 2.0) -- (0.5, 2.0);
\draw[->, red, thick] (0.5, -2.0) -- (-0.5, -2.0);
%\draw[->, red, thick] (-0.5 + 4, 2.0) -- (0.5, 2.0);
\draw[->, red, thick] (1.0 + 4.0, -2.0) -- (-1.0 + 4, -2.0);
\draw[->, red, thick] (0.5 + 8, 2.0) -- (-0.5 + 8, 2.0);
\draw[->, red, thick] (1.5 + 8, -2.0) -- (-1.5 + 8, -2.0);
% lambda
\node[] at (0,-2.5) {$\lambda_x = 0$};
\node[] at (4.0,-2.5) {$\lambda_x = \alpha \dot{\gamma} R$};
\node[] at (8.0,-2.5) {$\lambda_x = 2\alpha \dot{\gamma} R$};
% abc
\node[] at (-1.0,-2.5) {a)};
\node[] at (2.5,-2.5) {b)};
\node[] at (6.5,-2.5) {c)};
\end{tikzpicture}
\caption{\label{fig:lift_force_slip_velocity} Schematic of an intruder for three different slip velocities. The red arrows indicate the relative velocity of the top and bottom layer to the intruder, $\bm{x}$ represents an approximate force application point, $\bm{e}$ represent the angle of the drag force and $d_l$ is the diameter of a flow layer. a) There is no slip velocity and the forces are symmetric. b) The slip velocity equals the velocity of the top layer and c) The slip velocity is larger than the top layer. }
\end{center}
\end{figure*}
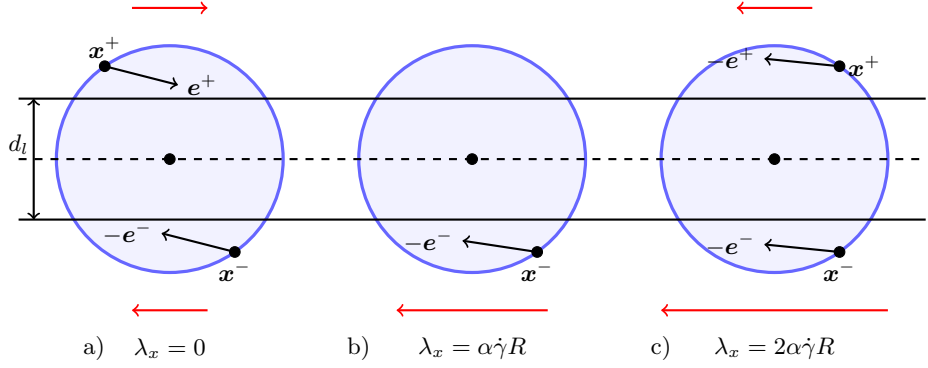

In Fig.~\ref{fig:lift_forces} the measured lift forces from Ref.~\cite{he2025lift} for various cases are shown as function of $\bar{\lambda}$. The overal trend shows an increase in lift force up till $\bar{\lambda_x} = 1$, afterwards it decreases and switches sign at $\bar{\lambda} \approx 3.5$.
With the force model in mind, this behaviour can now be explained. In Fig.~\ref{fig:lift_force_slip_velocity} the two drag forces are drawn schematically on an intruder. The angles of the drag forces in the schematic correspond to the measured $\theta^+$ and the red arrow indicates the velocity of the layer w.r.t the intruder. For a zero slip velocity, Fig.~\ref{fig:lift_force_slip_velocity}a), the flow is symmetric so the forces balance and no lift or drag is generated. The flow becomes asymmetric when a positive slip velocity is imposed, with the top layer having a lower relative velocity compared to the bottom layer, generating a force in the vertical direction. When the characteristic slip velocity $\lambda_x^+$ of the top layer (see Eq.~\eqref{eq:drag_force_top_slip_velocity}) becomes zero, the drag force at the top vanishes. This happens when the slip velocity equals the velocity of the top layer, $\lambda_x = \alpha \dot{\gamma}R$, which corresponds to the maximum lift measured on the intruder. The measure data shows that this peak occurs at $\bar{\lambda}_x = 1$ suggesting that $\alpha = 1 + S^{-1}$. In Fig.~\ref{fig:lift_force_slip_velocity}c) the top layer drag force will switch direction as the intruder now moves faster than the flow in the top layer.

\subsection{Remarks}
It has been shown that the force model is able to capture the measured $F_L$ and $\tilde{F}_d$. Given the measured angles of the competing drag forces in the top and bottom layer, the peak and reduction of the lift force can be explained. However, it is not yet clear what equation governs the drag force angles and this is closely related to the closure problems of section~\ref{sec:mododisperse_flows}. 

A worthy observation is that by actively breaking the symmetry of a mono-disperse flow by adding an intruder with slip velocity, it becomes possible to measure $\theta^+$ at zero slip velocity. This tool allows further study of the behaviour of the drag force angles for different (linear) mono-disperse flow configurations. 

This section shows that the use of the term lift force might be open for discussion as it actually seems to be caused by two tilted drag forces. Hence the proposed force model does not include a lift force explicitly.

\section{Buoyancy forces}
\label{sec:buoyancy_forces}
There is clear consensus in the literature that there is a granular buoyancy force acting on the intruder and that it scales with the pressure gradient. However, the exact definition is not agreed upon. There have been four proposed candidates:
\begin{enumerate}
\item All $\dz{p}$ related forces are attributed to the buoyancy force~\cite{jing2020rising}.
\item The component of the segregation force that does not scale with slip velocity~\cite{yennemadi2023drag}.
\item The buoyancy force follows an effective Voronoi volume~\cite{PhysRevFluids.3.074303,kumar2019theoretical}
\item The buoyancy force follows a reduction in surface contact density~\cite{lantman2021granular}
\end{enumerate}
The reason for this wide variation in definition is that not only the buoyancy force depends on the pressure gradient, but the effective drag force, scaling with $\dot{\gamma}\dz{\eta}$, does too and hence it is very hard to distinguish which part of the pressure gradient belongs to the buoyancy force and which to the effective drag force. This observation directly invalidates model 1. Model 2 would attribute the effective drag force to the buoyancy force as it does not directly depend on a slip velocity. Both model 3 and 4 assume that the buoyancy force matches the gravity force at $S=1$ which does not match with the observations from the flow equation in section~\ref{sec:mododisperse_flows}, where the buoyancy is required to be smaller. Hence, the only plausible buoyancy force model currently is the classical Archimedes buoyancy force.

\section{Segregation in non-linear flows}
\label{sec:intruder_mechanics}
This section focusses on the behaviour of intruders in non-linear granular flows. Before diving into the various aspects of the force model, insights from mono-disperse results in section~\ref{sec:mododisperse_flows} allows for simplifications of Eq.~\eqref{eq:total_force_model}. In non-linear flows the term $\tau\alpha R^2$ can be rewritten using Eq.~\eqref{eq:orthogonal_balance},
\begin{equation}
\tau\alpha R = \dot{\gamma} \dz{\eta} \alpha^2 R^2 \psi_f,
\end{equation}
where $\psi_f$ is Eq.~\eqref{eq:psi} for the reference flow which is in the same form as the effective drag force. Similarly, $\dz{\eta}$ can be eliminated from $\lambda_x \dz{\eta}\alpha R$ by rewriting Eq.~\eqref{eq:orthogonal_balance} as
\begin{equation}
\dz{\eta} = \frac{\eta}{\alpha \psi_f R}
\end{equation}
such that
\begin{equation}
\lambda_x \dz{\eta}\alpha R =  \eta \lambda_x \frac{1}{\psi_f}
\end{equation}
becomes a Stokesian drag force. The full force model can thus be written as
\begin{equation}
\label{eq:Intruder_force_model}
\bm{F}_e + (D - \phi) \rho_f \bm{g} V_I - c \eta 
\begin{bmatrix}
\bm{l} & \bm{e}_z
\end{bmatrix}
\bm{\lambda} R 
+ \frac{3}{4 \pi} c\alpha^2\dot{\gamma}\dz{\eta} V_I \bm{k} = \bm{0}
\end{equation}
where the vector $\bm{l}$ is
\begin{equation}
\bm{l} = \left[\bm{e}^{+} + \bm{e}^{-} \right] + \frac{1}{\psi_f} \left[\bm{e}^{+} - \bm{e}^{-}\right].
\end{equation}
Note that the intruder experiences once again a drag force with two different contributions, a Stokes-like drag force and the effective drag force due to a difference in magnitude in the top and bottom region.

\subsection{Segregation force}
\label{sec:segregation_force}
In the pursuit of understanding segregation, simulations with an intruder-on-a-spring method are often employed. In this method the intruder is fixed to a spring in the $z$-direction, taking away the slip velocity in that direction. The general equation is~\cite{guillard2016scaling,jing2021unified}
\begin{equation}
\label{eq:segregation_force_definition}
F_s + F_{g_z} + F_{\textrm{seg}} = 0,
\end{equation}
where $F_s$ is the spring force, $F_{g_z}$ the gravity force and $F_{\textrm{seg}}$ the segregation force. The segregation force for the current model is, see Appendix~\ref{sec:appendix_intruder},
\begin{equation}
\label{eq:seg_force}
F_{\textrm{seg,m}} \equiv \left(
\mathcal{H}_0 \dz{\tau} 
+ \mathcal{H}_1 \dot{\gamma}\dz{\eta}
- \dz{p}
 \right) V_I,
\end{equation}
where the subscript $m$ marks that this is the segregation force definition of the force model considered in this work. The first term is related to the horizontal slip velocity with 
\begin{equation}
\mathcal{H}_0 = \frac{l_z}{l_x}(1 / \phi - 1)
\end{equation}
The second term is caused by the effective drag force with
\begin{equation}
\mathcal{H}_1 = \frac{3}{4 \pi} c \alpha^2 \left(k_z - \frac{l_z}{l_x} k_x \right),
\end{equation}
where the latter term comes from the Stokesian drag component in the $x$-direction. The last term is the Archimedes buoyancy force. This segregation force is compared to two unified segregation force models in the next section.

\subsection{Unified segregation force models}
\label{sec:unified_segregation_force}
Two unified segregation force models have been proposed. First consider the scaling laws from~\cite{guillard2016scaling} which were obtained for two dimensional simulations,
\begin{equation}
\label{eq:guillard}
F_{\textrm{seg,g}} = -\left( \mathcal{F}(\mu, S) \dz{p} + \mathcal{G}(\mu, S) \dz{\tau} \right) V_I.
\end{equation}
Two scaling laws were found for two distinctive cases. The first case showed that $\left. F_{\textrm{seg,g}}\right|_{\dz{p} = 0} \propto \dz{\tau}$. The second case showed that $\left. F_{\textrm{seg,g}}\right|_{\dz{\tau} = 0} \propto \dz{p}$. It was then shown that these two effects are additive. It is assumed that the additive property of $\dz{p}$ and $\dz{\tau}$ extend to three dimensions with spheres, but the scaling functions $\mathcal{F}$ and $\mathcal{G}$ may be different. 

This scaling law agrees with the currently proposed segregation force as follows. Take the dot product between $\bm{k}$ and Eq.~\eqref{eq:flow_equation_vector_form},
\begin{equation}
\dot{\gamma}\dz{\eta} = \frac{4 \pi}{3} \frac{1 - 1 / \phi}{c \alpha^2}\phi \rho_f (g_x k_x + g_z k_z).
\end{equation}
Next, substitute Eq.~\eqref{eq:dztau} and Eq.~\eqref{eq:dpdz},
\begin{equation}
\dot{\gamma}\dz{\eta} = \frac{4 \pi}{3} \frac{1 - 1 / \phi}{c \alpha^2} \left(-\dz{\tau}k_x + \dz{p} k_z \right).
\end{equation}
Substituting this formulation back into Eq.~\eqref{eq:seg_force} leads to
\begin{equation}
\label{eq:fseg_as_guillard}
F_{\textrm{seg,m}} = \left( \mathcal{A}_p \dz{p} + \mathcal{A}_{\tau} \dz{\tau} \right) V_I,
\end{equation}
with
\begin{equation}
\mathcal{A}_p = \frac{4 \pi} {3} \frac{(1 - 1 / \phi)}{c \alpha^2}  k_z \mathcal{H}_1 - 1,
\end{equation}
\begin{equation}
\mathcal{A}_{\tau} = \mathcal{H}_0 - \frac{4 \pi} {3} \frac{(1 - 1 / \phi)}{c \alpha^2} k_x \mathcal{H}_1.
\end{equation}
This shows that both models describe the same scaling of $F_{\textrm{seg}}$ with pressure and shear stress gradients. Hence, it follows that $\mathcal{A}_p = - \mathcal{F}(\mu,S)$ and $\mathcal{A}_{\tau} = - \mathcal{G}(\mu,S)$. The observation that the pressure- and shear-stress gradients are additive follows naturally from the force model in this work, since $\dot{\gamma} \dz{\eta}$ scales additively with $\dz{p}$ and $\dz{\tau}$, compensating the gap between the gravity force and the buoyancy force.

The second unified segregation force model was proposed in~\cite{jing2021unified} where it was stated that
\begin{equation}
\label{eq:ling}
F_{\textrm{seg,l}} = \left(- h^g(S) \dz{p} + h^k(S) \frac{p}{\dot{\gamma}}\dz{\dot{\gamma}}\right) V_I.
\end{equation}
Here it was reasoned that the first term represents a gravity induced contribution the segregation force, while the latter is a kinematic component. The argument to propose a different scaling law than Eq.~\eqref{eq:guillard} was that the scaling functions $\mathcal{F}$ and $\mathcal{G}$ depend on $\mu$ while $p$ and $\tau = \mu p$ are also related, by definition. It was suggested that the relations proposed by Eq.~\eqref{eq:ling} should therefore only depend on $S$. 

To compare $F_{\textrm{seg,l}}$ with $F_{\textrm{seg,m}}$, expand $\dz{\tau} = \dz{\eta \dot{\gamma}} = \eta \dz{\dot{\gamma}}+ \dot{\gamma} \dz{\eta}$ and expand $\dot{\gamma}\dz{\eta}$ with respect to $\dz{p}$ and $\frac{p}{\dot{\gamma}}\dz{\dot{\gamma}}$ using Eq.~\eqref{eq:deta_derivative}, leading to
\begin{equation}
\label{eq:fseg_as_ling}
F_{\textrm{seg}, m} = \left(-\mathcal{B}_p \dz{p} + \mathcal{B}_{\tau} \frac{p}{\dot{\gamma}}\dz{\dot{\gamma}}\right) V_I,
\end{equation}
with
\begin{equation}
\label{eq:seg_func_g}
\mathcal{B}_p = 1 - \left( \mu - \frac{I}{2}\frac{\partial \mu}{\partial I}  \right) (\mathcal{H}_0 + \mathcal{H}_1) = h^g,
\end{equation}
\begin{equation}
\label{eq:seg_func_k}
\mathcal{B}_{\tau} =  \mu \mathcal{H}_0 -  \left(\mu - I\frac{\partial \mu}{\partial I} \right)  (\mathcal{H}_0 + \mathcal{H}_1) = h^k.
\end{equation}
Observe that indeed $F_{\textrm{seg}, m} $ can be cast into the same scaling as $F_{\textrm{seg,l}}$, although there is a clear dependency on the $\mu(I)$-rheology. One explanation could be that the forces measured in~\cite{jing2021unified} were in a range of $I \in [0.05,0.44]$ while the functions $\mathcal{F}$ and $\mathcal{G}$ seem to vary with friction mostly in the region of $I \in [0.005, 0.05]$. It might therefore be possible that the friction dependence might have been missed. 

Since $F_{\textrm{seg,m}}$ can be cast in both the form of $F_{\textrm{seg,g}} $ and $F_{\textrm{seg,l}}$ it has been shown that the two unifying segregation force models actually describe the same effect, albeit with a different explanation. All three models agree that there are two competing components, but disagree on what competes. 
Since the competing drag force theory is also able to reproduce velocity profiles for mono-disperse flows, it follows that this is the most natural explanation. For practical purposes, the user can choose which form seems to be the most suitable for the required application.

\subsection{Segregation mechanism}
The origin of segregation in dense granular flows lies in the delicate balance between the buoyancy and drag forces with the gravity force on a particle. The model suggests that the mechanism of segregation is always present in dense granular flows and that only mono-disperse flow is a very special case where the buoyancy and drag forces exactly balance the gravity force. As soon as an intruder changes properties, such as shape, density or friction, it will start to segregate.

\begin{figure}
\begin{center}
\includegraphics[width=0.99\linewidth]{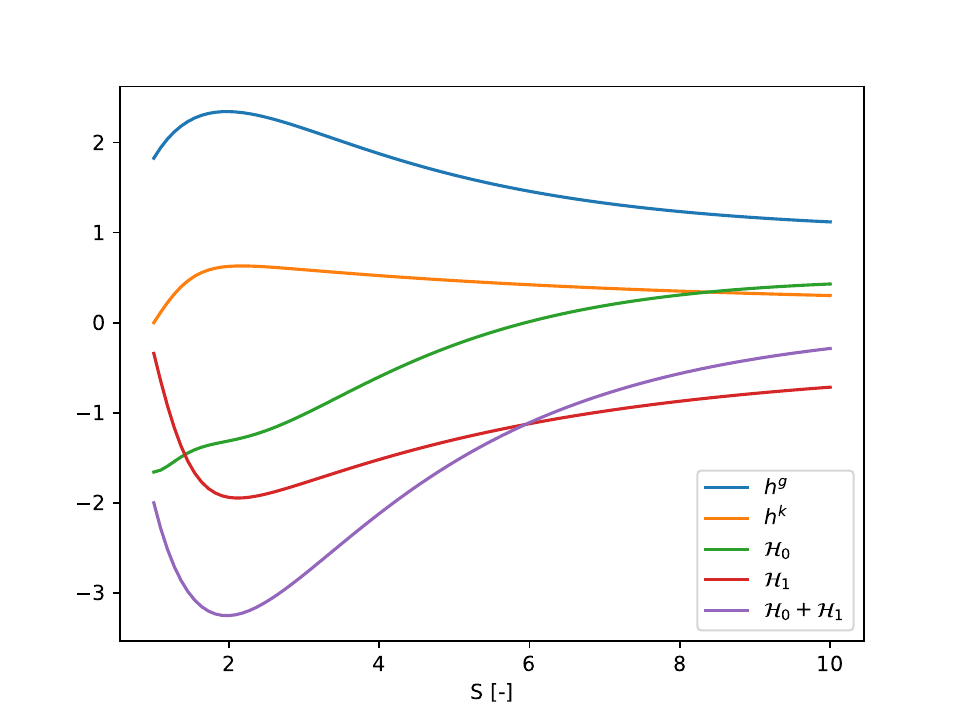}
\caption{\label{fig:scaling_functions} This graph shows the emperical fit functions $h^g = (1 - 1.34 e^{-1.09 S}) (1 + 3.55 e^{-0.34S})$ and $h^k = 0.19 \tanh(1.69[S - 1])(1 + 3.63 e^{-0.18S})$ from Ref.~\cite{jing2021unified}. The functions $\mathcal{H}_0$ and $\mathcal{H}_1$ are computed using Eq.~\eqref{eq:seg_func_g} and Eq.~\eqref{eq:seg_func_k} with $\mu(I) = 0.36 + 0.58 /(0.8 / I + 1)$ and $I=0.15$.}
\end{center}
\end{figure}

An important question for size segregation is the segregation direction. For this purpose, slightly rewrite Eq~\eqref{eq:seg_force} by expanding $\dz{\tau} = \eta \dz{\dot{\gamma}} + \dot{\gamma} \dz{\eta}$,
\begin{equation}
F_{\textrm{seg,m}} \equiv \left(
\mathcal{H}_0 \eta \dz{\dot{\gamma}} 
+ (\mathcal{H}_0 + \mathcal{H}_1) \dot{\gamma}\dz{\eta}
- \dz{p}
 \right) V_I.
\end{equation}
The functions $\mathcal{H}_0$ and $\mathcal{H}_1$ can be reconstructed from Eq~\eqref{eq:seg_func_g} and Eq.~\eqref{eq:seg_func_k}, combined with the fit functions of $h^g$ and $h^k$ from~\cite{jing2021unified} and are shown in Fig.~\ref{fig:scaling_functions}. Both $\mathcal{H}_0$ and $\mathcal{H}_1$ are negative at $S=1$. The first function slowly increases as $S$ increases and becomes positive at $S > 6$. The second shows a strong increase in magnitude till roughly $S \approx 2$ and then slowly decreases again afterwards. At $S=1$ the segregation force matches the gravity force in magnitude, but as $S$ increases the middle term, scaling with $\dot{\gamma}\dz{\eta}$ starts to dominate as it has the strongest dependency on $S$. Since $\mathcal{H}_0 + \mathcal{H}_1$ is negative, it follows that the intruders with roughly $S=2$ segregate towards lower viscous regions. This fundamental result has been previously observed~\cite{thesisMarnix}. For size ratios of $S > 6$ the magnitude of the first and second term reduce significantly, which may lead to so called reverse segregation~\cite{gamble2026reversed, thomas2018evidence}.

For an analysis of the angles of the drag forces on the intruder functions $\mathcal{H}_0$ and $\mathcal{H}_1$ are not sufficient as there are three unknowns to find, $\theta^+$, $\theta^-$ and $\psi_f$. Future numerical investigations may be able to overcome this limitation by also measuring the slip velocity of the intruder or by reconstructing the velocity profile using Eq.~\eqref{eq:flow_equation}.

\section{Discussion}
\label{sec:discussion}
The current work only considers a single intruder in an otherwise uniform bulk flow. However, this case is mostly interesting for academic purposes to understand the fundamental behaviour of segregation. In real applications the flow consists of many particles with different sizes an densities which will have more complex interactions. The force model may be extended by having the coefficients such as $\alpha$ and $c$ depend on the particle properties locally to the intruder. This can be done by employing an extension of the particle-on-a-spring method to flows with various intruder concentrations~\cite{Duan_Jing_Umbanhowar_Ottino_Lueptow_2022,Duan_Jing_Umbanhowar_Ottino_Lueptow_2024}. 

Another assumption of the force model is that the flow consists of straight layers where the asymmetric drag force act parallel to the layers with a slight orthogonal angle. This method could in principle be extended to three dimensional flows in more complex geometries, however it is unclear how the drag forces behave when the layers actually bend and warrants further investigation.

While in this work it is assumed the intruder does not rotate, it could be included by extending the slip velocity definitions $\lambda^+$ and $\lambda^-$ with an angular velocity correction.

The force model will be valid for intruders with ratios slightly smaller that one, however, it will transition into a percolation mechanism~\cite{jain2005regimes} as small intruders don't fully partake in the fluid and could reside within a flow layer.

\section{Conclusion}
\label{sec:conclusions}
A force model for an intruder in a mono-disperse dense granular flow is derived by identifying that there are two opposing drag forces acting on the intruder. The critical observation here is that the drag caused by the horizontal lag velocity may generate a vertical force component. The model was then complemented with gravity, buoyancy and external forces. The two opposing drag forces induce a new drag term that scales with the shear rate, viscosity gradient and the volume of the intruder.

A key insight is that the force model must also hold for all particles in mono-disperse flows, leading to a flow equation that determines a family of velocity profiles for dense granular flows, including a Bagnold velocity profile. Interestingly, the general structure of the velocity profile is governed by the gravity force, but the specific flow profile follows from a micro-mechanical balance that remains unclear.

The force model reveals insight to long standing confusion in the field. There has been discussion on what the  buoyancy force should be on an intruder in a dense granular flow. This force model suggests that the buoyancy force is simply the classical Archimedes buoyancy force. In addition, it appears that lift forces are actually the orthogonal component of drag forces and bear no direct equivalent force in classical fluids. The force model can be cast into two different unified segregation models together and show that particles with moderate size ratio migrate to lower viscous regions.

Perhaps the most remarkable result is that this work shows how tightly coupled dense granular flow rheology and segregation are as a single force model has the capability to describe both phenomena. It therefore opens a totally new perspective on dense granular flows. There are many future topics to investigate. One research direction is to investigating $\bm{e}^+$, $\bm{e}^-$ and $\psi$ in mono-disperse flows to measure and understand what equation is governing these quantities with the ultimate goal of understanding the $\mu(I)$-rheology. Another direction is to extend the validity of the current force model by including curved stream-lines, and by extending it to multiple intruders.

\section{Acknowledgements} 
The author is grateful for the feedback and discussions with Kasper van der Vaart and to Richard Lueptow for feedback and sharing the data from Ref.~\cite{he2025lift}.
\appendix

\section{Flow solution derivations}
\label{sec::appendix_a}
Here the solutions to Eq.~\eqref{eq:flow_equation} are derived. A more convenient form for solving this ODE is obtained by expanding the flow equation using the chain rule using $\eta = \tau / \dot{\gamma}$,
\begin{equation}
\label{eq:flow_equation_2}
\frac{1}{\dot{\gamma}} \dz{\dot{\gamma}} = \frac{1}{\tau} \left[ \dz{\tau} + \rho_f|\bm{g}|(\phi - 1)f_c \right].
\end{equation}
There are three distinct cases, depending on $\bm{g}$.

\subsection{Zero gravity} In absence of gravity, $|\bm{g}| = 0$, it follows that there is no shear stress gradient, see Eq.~\eqref{eq:dztau}. Therefore the whole right-hand-side of Eq.~\eqref{eq:flow_equation_2} will vanishes, leading to
\begin{equation}
\frac{1}{\dot{\gamma}} \dz{\dot{\gamma}} = 0,
\end{equation}
It follows that the viscosity must be constant,
\begin{equation}
\dot{\gamma} = A_1 \tau_0.
\end{equation}
Integrating over $z$ and enforcing the boundary conditions $v(0) = 0$ and $v(h) = v_h$ give a linear profile,
\begin{equation}
v(z) = v_0 + \frac{v_h - v_0}{h} z.
\end{equation}

\subsection{Vertical gravity} When $g_x = 0$ and $|g_z| > 0$, the shear stress is constant and so Eq.~\eqref{eq:flow_equation_2} becomes
\begin{equation}
\frac{1}{\dot{\gamma}} \dz{\dot{\gamma}} = \frac{\rho_p |g_z|(\phi - 1) f_c}{\tau_0}.
\end{equation}
This equation can be solved using separation of variables,
\begin{equation}
\dot{\gamma} = B_1e^{B_2 z}, \quad B_2 = \frac{\rho_p|g_z| (\phi - 1) f_c}{\tau_0}
\end{equation}
where $B_1$ is an integration constant. Next, integrate over $z$ and apply the boundary conditions the obtain the velocity profile,
\begin{equation}
v = v_0 + (v_h - v_0) \frac{1 - e^{B_2 z}}{1 - e^{B_2 h}}.
\end{equation}

\subsection{Vertical and horizontal gravity} In this case substitute the linear shear stress profile $\tau = \tau_0 + \dz{\tau} z$ into Eq.~\eqref{eq:flow_equation_2}, 
\begin{equation}
\frac{1}{\dot{\gamma}} \dz{\dot{\gamma}} = \frac{1}{\tau_0 + \dz{\tau} z} \left[ \dz{\tau} + \rho_f|\bm{g}|(\phi - 1) f_c \right].
\end{equation}
The equation can be rewritten as
\begin{equation}
\frac{1}{\dot{\gamma}} \dz{\dot{\gamma}} = \frac{1}{\tau_0 / \dz{\tau} + z} \left[1 + \frac{\rho_f|\bm{g}|(\phi - 1) f_c}{\dz{\tau}} \right].
\end{equation}
After substitution of Eq.~\eqref{eq:dztau} and defining $\bar{\tau} = \tau_0 / (\phi \rho_f g_x)$ the equation becomes
\begin{equation}
\frac{1}{\dot{\gamma}} \dz{\dot{\gamma}} = \frac{1}{-\bar{\tau} + z} \left[1 - \frac{|\bm{g}|(1 - 1 / \phi) f_c}{g_x} \right].
\end{equation}
A more convenient form is
\begin{equation}
\frac{1}{\dot{\gamma}} \dz{\dot{\gamma}} = \frac{C_1 - 1}{\bar{\tau} - z} 
\end{equation}
with 
\begin{equation}
C_1 = \frac{|\bm{g}|}{g_x} (1 - 1 / \phi) f_c
\end{equation}
This equation can be solved by separation of variables,
\begin{equation}
\ln \dot{\gamma} = \ln \left( \left (\bar{\tau} - z \right)^{C_1 - 1}\right) + C_2.
\end{equation}
with $C_2$ some integration constant. Taking the exponential on both sides, the equation for shear rate is obtained,
\begin{equation}
\dot{\gamma} = C_3(\bar{\tau} - z)^{C_1 - 1}
\end{equation}
with $C_3 = e^{C_2}$. Integration over $z$ gives
\begin{equation}
v(z) = C_4 + C_3 \frac{\left(\bar{\tau} - z \right)^{C_1}}{C_1}
\end{equation}
with $C_4$ another integration constant. After substitution of the boundary conditions the final form is
\begin{equation}
\label{eq:power_equation}
v(z) = v_0 + (v_h - v_0) \frac{(\bar{\tau} - z)^{C_1} - \bar{\tau}^{C_1}}{(\bar{\tau} - h)^{C_1} - \bar{\tau}^{C_1}}.
\end{equation}

\section{Segregation force}
\label{sec:appendix_intruder}
To obtain an expression for the segregation force $F_{\textrm{seg}}$, defined by Eq.~\eqref{eq:segregation_force_definition},  
\begin{equation}
F_s + F_{g_z} + F_{seg} = 0,
\end{equation}
consider the $z$-direction of Eq.~\eqref{eq:Intruder_force_model} with $F_s$ as a vertical external force, $\rho_f g_z V_I$ replaced with $F_{g_z}$ and the spring force is designed such that no vertical motion is possible, $\lambda_z = 0$, leading to
\begin{equation}
\label{eq:size_ratio_balance_z}
F_s + F_{g_z}
- \phi \rho_f g_z V_I 
- c \eta l_z \lambda_x R
+ \frac{3}{4 \pi} c\dot{\gamma}\dz{\eta} V_I k_z,
 = 0.
\end{equation}
The $x$-direction of Eq.~\eqref{eq:Intruder_force_model} is
\begin{equation}
(1 - \phi) \rho_f g_x V_I - c \eta l_x \lambda_x R + \frac{3}{4\pi} c \alpha^2 \dot{\gamma} \dz{\eta} V_I k_x = 0,
\end{equation}
which can be reordered as
\begin{equation}
c \eta l_z \lambda_x R = \frac{l_z}{l_x} \left( 
(1 - \phi) \rho_f g_x V_I + \frac{3}{4\pi} c \alpha^2 \dot{\gamma} \dz{\eta} V_I k_x
\right).
\end{equation}
Substitute this back into Eq.~\eqref{eq:size_ratio_balance_z} to eliminate the slip velocity,
\begin{equation}
F_s + F_{g_z}
- \phi \rho_f g_z V_I 
- \mathcal{H}_0 \phi \rho_f g_x V_I
+ \mathcal{H}_1 \dot{\gamma}\dz{\eta} V_I = 0,
\end{equation}
where
\begin{equation}
\mathcal{H}_0 = \frac{l_z}{l_x}(1 / \phi - 1)
\end{equation}
and
\begin{equation}
\mathcal{H}_1 = \frac{3}{4 \pi} c \alpha^2 \left(k_z - \frac{l_z}{l_x} k_x \right)
\end{equation}
Finally, substitute Eq.~\eqref{eq:dpdz} and Eq.~\eqref{eq:dztau},
\begin{equation}
F_s + F_{g_z}
+ \left(\mathcal{H}_0 \dz{\tau}
+ \mathcal{H}_1 \dot{\gamma}\dz{\eta}
- \dz{p}
 \right) V_I
 = 0.
\end{equation}
The segregation force of equivalent of the force model can therefore be expressed as
\begin{equation}
\label{eq:seg_force_derived}
F_{\textrm{seg,m}} \equiv \left(
\mathcal{H}_0 \dz{\tau} 
+ \mathcal{H}_1 \dot{\gamma}\dz{\eta}
- \dz{p}
 \right) V_I.
\end{equation}

\section{Derivative of the viscosity gradient}
\label{sec:appendix_b}
This appendix contains the derivation of the relation between $\dot{\gamma} \dz{\eta}$ and the two gradients $\dz{p}$ and $\dz{\dot{\gamma}}$. For simplicity only positive shear rate is assumed. The viscosity in a $\mu(I)$-granular flow can be expressed as
\begin{equation}
\eta = \frac{\mu p}{\dot{\gamma}},
\end{equation}
With 
\begin{equation}
\mu = \mu_1 + \frac{\mu_2 - \mu_1}{I_0 / I + 1}
\end{equation}
and
\begin{equation}
I = \frac{d \dot{\gamma}}{\sqrt{p / \rho}}.
\end{equation}
The derivative of the viscosity with respect to $z$ is
\begin{equation}
\dz{\eta} =
\frac{\mu}{\dot{\gamma}} \dz{p}
+ \frac{p}{\dot{\gamma}} \frac{\partial \mu}{\partial I} \dz{I}
- \frac{\mu p}{\dot{\gamma}^2} \dz{\dot{\gamma}}.
\end{equation}
The $z$-derivative of the inertial number is
\begin{equation}
\label{eq:dIdz}
\dz{I} = 
\frac{I}{\dot{\gamma}}\dz{\dot{\gamma}}
- \frac{I}{2p} \dz{p}
\end{equation}
and so the derivative of viscosity can be expressed as
\begin{equation}
\dz{\eta} = \left(
\frac{\mu}{\dot{\gamma}} -
\frac{1}{\dot{\gamma}} \frac{\partial \mu}{\partial I} \frac{I}{2}
\right) \dz{p}
+ 
\left(
I \frac{\partial \mu}{\partial I} \frac{p}{\dot{\gamma}^2}
- \frac{\mu p}{\dot{\gamma}^2}
\right) \dz{\dot{\gamma}}.
\end{equation}
Multiplication with $\dot{\gamma}$ and some rearranging gives
\begin{equation}
\label{eq:deta_derivative}
\dot{\gamma}\dz{\eta} = \left( \mu - \frac{I}{2}\frac{\partial \mu}{\partial I}  \right) \dz{p}
- \left(\mu - I\frac{\partial \mu}{\partial I} \right) \frac{p}{\dot{\gamma}} \dz{\dot{\gamma}}.
\end{equation}
\bibliography{main}{}
\bibliographystyle{plain}

\end{document}